\documentclass[12pt]{iopart}
\usepackage{graphicx}

\newcommand{\newc}{\newcommand}
\newc{\beq}{\begin{equation}}
\newc{\enq}{\end{equation}}
\newc{\bea}{\begin{eqnarray}}
\newc{\ena}{\end{eqnarray}}
\newc{\D}{\displaystyle}
\newc{\rt}{\right}
\newc{\lt}{\left}
\newc{\bfk}{{\mathbf k}}
\newc{\bfq}{{\mathbf q}}
\newc{\bfp}{{\mathbf p}}
\newc{\poi}{\phantom . \noi}
\newc{\nn}{\nonumber}
\newc{\sqp}{\phi}
\newc{\cp}{\varphi}

\begin{document}

\title[Inflation and nonequilibrium renormalization group]{Inflation and nonequilibrium renormalization group}

\author{Juan Zanella and Esteban Calzetta}

\address{CONICET and Departamento de F\'{\i}sica, Universidad de Buenos Aires, Ciudad Universitaria, 1428
Buenos Aires, Argentina} \ead{zanellaj@df.uba.ar,
calzetta@df.uba.ar}
\begin{abstract}
We study de spectrum of primordial fluctuations and the scale
dependence of the inflaton spectral index due to self-interactions
of the field.  We compute the spectrum of fluctuations by applying
nonequilibrium renormalization group techniques.
\end{abstract}

\submitto{\JPA}

In this paper we explore a mechanism to explain departures of the
primordial fluctuations from the spectrum of a free inflaton
field. This is important, because the spectrum of the inflaton
field at the exit of the horizon is directly related with the
level of inhomogeneity of the observed universe. The question we
want to address is how the nonlinearities for an interacting field
affect the predicted spectrum \cite{Calzetta:Hu,
Calzetta:Gonorazky, Boyanovsky:Vega, chinos}. To this end we will
use a nonequilibrium renormalization group (RG).

If the inflaton field were a free field, its spectrum would be of
Harrison-Zel'dovich type. At the horizon exit (HE) of the mode
with wave number $k$ it would be \bea  \lt<\Phi(k,t)
\Phi(k,t)\rt>_{\rm HE} \propto \D\frac{1}{k^3}. \label{HZ} \ena
The spectral index $n(k)$ measures deviations from this law. It is
defined by \bea \label{spectral index} \lt<\Phi(k,t)
\Phi(k,t)\rt>_{\rm HE} \propto \D\frac{1}{k^3} \; k^{n(k)-1}. \ena
Hence, the Harrison-Zel'dovich spectrum (\ref{HZ})  has $n(k) =
1$. Present experiments show that the spectral index is close to
one and that presumably it runs with the scale, \bea n(k) = 1 +
\Delta(k), \ena just at the edge of the experimental precision
\cite{Casas}.

We will compute the spectrum of the inflaton field  for a toy
model of inflation. The goal is to show how a RG defined for
nonequilibrium problems can be used to predict the spectrum of an
interacting inflaton field.  We will assume a spatially flat
Robertson-Walker metric with constant expansion rate $H$, \bea
ds^2 = dt^2 - a^2(t) \lt(dx^2 + dy^2 + dz^2\rt), \;\;\;\; a(t) =
a_0 e^{H t},\ena and an inflaton field described by a massless
$\lambda \phi^4$ action \bea \fl S[\Phi] = \D\int \!dt \; a(t)^3
\Bigg[&\D\int d^d\!q\; \frac{1}{2} \lt(\dot \Phi^2 - q^2
\D\frac{\Phi^2}{a(t)^2}+ \xi  R \;\Phi^2 \rt) \nn \\ &-
\D\frac{\lambda}{4!} \D\int \prod_{i=1}^4 d^d\!q_i \;\;
\delta^3\!\lt(\bfq_1+\dots+\bfq_4\rt) \;\, \Phi_1 \Phi_2 \Phi_3
\Phi_4 \Bigg]. \ena [$\Phi^2 \equiv \Phi(\bfq, t) \Phi(-\bfq, t)$,
and $\Phi_i \equiv \Phi(\bfq_i, t)$.] Transforming to conformal
coordinates \numparts \bea \eta = -({a H})^{-1},  \\ a \Phi =
\Phi_c, \label{conformal relation}\ena \endnumparts and ignoring
all the mass terms, the original theory is mapped into a scalar
field theory in flat space-time with time coordinate $-\infty <
\eta < 0$,  \bea \label{conformal action} \fl S[\Phi_c] = \D\int
\!d\eta \Bigg[\D\int d^d\!q\; \frac{1}{2} \lt(\Phi_c^{'2} - q^2 \,
\Phi_c^2\rt) - \D\frac{\lambda}{4!} \D\int \prod_{i=1}^4 d^d\!q_i
\;\; \delta^3\!\lt(\bfq_1+\dots+\bfq_4\rt) \;\, \Phi_{c1}
\Phi_{c2} \Phi_{c3} \Phi_{c4} \Bigg], \ena where $\Phi_c' =
\partial \Phi_c/\partial \eta$. We can show that
if $\lambda = 0$ the expression (\ref{espectro 0}) is regained.
Note that when $\lambda = 0$, the expectation value for the
product of two conformal fields $\Phi_c$ is given by the usual
expression for a free field in flat space-time \bea
\lt<\Phi_c(k,\eta) \Phi_c(k, \eta)\rt> \propto \D\frac{1}{k} .\ena
Using (\ref{conformal relation}), the expectation value for the
original theory is \bea \lt<\Phi(k, t) \Phi(k,t)\rt> \propto
\D\frac{1}{a^2(t) k}. \label{espectro 0} \ena To obtain the
spectrum, this expression must be evaluated when the mode $k$
exits the horizon. This happens when its physical wavelength
$k^{-1} a(t)$ equals the horizon size $H^{-1}$, that is, when \bea
a(t) = k H^{-1}.\label{eta exit} \ena Thus, replacing this value
of $a$ in (\ref{espectro 0}), Eq. (\ref{HZ}) is recovered. We will
use RG techniques to compute the spectrum when $\lambda \ne 0$.

The basic idea of RG for systems in equilibrium (where time does
not enter in the description) is the coarse graining of the
original system,  i.e. the change in the resolution with which the
system is observed \cite{WiKo74}. Given a system with a range of
scales which goes up to wave number $\Lambda$,  if we are only
interested in scales up to wave number $k<\Lambda$, we can
separate the original system in two sectors: a lower wave number
sector, with $k'<k$, the relevant system, and a higher wave number
sector with $k< k'<\Lambda$,  the environment. Once this division
is done, the environment modes are eliminated from the
description. In equilibrium, this is achieved by computing the
coarse grained 'in-out' effective action for the lower sector,
complemented with a rescaling of the fields and momenta that
restores the cutoff and the coefficient of the $q^2$ term in the
action to their initial values. The elimination of the modes
between $\Lambda$ and $k$ proceeds by infinitesimal steps. In this
way, the calculation involves only tree and one loop diagrams, and
the resulting equations form a set of differential equations for
the parameters that define the effective action \cite{WH}.

Essentially, the same scheme can be used for nonequilibrium
systems. The main difference with the usual approach is that the
time variable must enter in the description. We want to compute
true expectation values at given times, not transition amplitudes
between 'in' and 'out' asymptotic states, far away in the future
and in the past. We want to follow the real and causal evolution
of expectation values, for which the usual 'in-out' representation
is not appropriate. A suitable description of nonequilibrium
systems is given within the 'closed time path' (CTP) formalism
\cite{ctp1, ctp2,ctp3, ctp4, dalvit:1996a}. The number of fields
is doubled, path integrals are over two histories $\cp^+$ and
$\cp^-$, that coincide at the time of observation $T$, which is a
new dimensional parameter of the theory.

The possibility of couplings between the two histories enlarge the
parameter space. Notably, it includes noise and dissipation.
Written in terms of $\sqp = \cp^+ - \cp^-$ and $\cp = \cp^+ +
\cp^-$, the free action for a scalar field in the CTP formalism is
\begin{eqnarray}   S_0[\phi, \varphi] =
\displaystyle\int_{0}^{T} \!\!dt \displaystyle\int\! d^d\!q
\;\bigg[ &\frac{1}{2} \dot\phi \; \dot\varphi -
\displaystyle\frac{1}{2} \left(q^2 + m^2\right) \phi\cp \label{the
action 0}  &- \kappa \; \phi \dot\varphi+ \displaystyle\frac{i}2
\nu \; \phi \phi \bigg]. \end{eqnarray} [$\sqp \cp \equiv
\sqp(\bfq, t) \cp(-\bfq, t)$, etc.] Here $\kappa$ is associated to
dissipation and $\nu$ to noise. For a given order $n$ in the
fields, there are $n$ possible interaction terms. Thus, for
example, the quartic interactions that can appear in the CTP
action are \bea \D\int_{0}^{T} \!\!\prod_{i=1}^4 dt_i \!\D\int
\!\!\prod_{i=1}^4 d^d\!q_i &\Big[v_{41}(q,t) \;\, \sqp_1 \; \cp_2
\cp_3 \cp_4 + i \, v_{42}(q,t) \;\, \sqp_1 \sqp_2 \; \cp_3 \cp_4
\nn \\ &+v_{43}(q,t) \;\, \sqp_1 \sqp_2 \sqp_3 \; \cp_4 + i \,
v_{44}(q,t) \;\, \sqp_1 \sqp_2 \sqp_3 \sqp_4 \; \Big]. \ena
[$\sqp_i = \sqp(q_i, t_i)$, etc.] In principle, all the allowed
couplings must be taken into account to compute the RG equations.

Even when the initial action at scale $\Lambda$ is the usual,
local, massless, $\lambda \cp^4$ action, which in the CTP
representation reads \bea \label{original theory} \fl S_\Lambda =
\D\int_{0}^{T} \!dt \lt[\D\int d^d\!q\; \frac{1}{2} \lt(\dot
\sqp\, \dot \cp - \, q^2 \sqp \,\cp \rt) - \D\frac{\lambda}{48}
\D\int \prod_{i=1}^4 d^d\!q_i \lt(\sqp_1 \; \cp_2 \cp_3 \cp_4 + \,
\sqp_1 \sqp_2 \sqp_3 \; \cp_4\rt) \rt], \ena as short wave number
modes are eliminated out, the RG flow generates all the possible
terms, with an arbitrary number of fields and nonlocal
dependencies (but with certain constrains imposed by the CTP
formalism). However, after a brief excursion, most of these terms
will go to zero along with $\lambda$, except for a few terms in
the free action. The effective action at scale $k$ will be given
approximately  by (\ref{the action 0}), but its parameters will
depend on both, the scale $k$ and the time $T$ \bea  \fl S_k \sim
\D\int_{0}^{T} \!\!dt \D\int d^d\!q\; \Bigg\{\D\frac{1}{2}\dot
\sqp\, \dot \cp &- \D\frac{1}{2}\bigg[q^2 +
\lt(\frac{\Lambda}{k}\rt)^2 m^2(k, T)\bigg]\sqp \,\cp \nn \\  &-
\lt(\frac{\Lambda}{k}\rt) \kappa(k, T) \, \sqp \,\dot \cp +
\D\frac{i}{2} \lt(\frac{\Lambda}{k}\rt)^2 \nu(k, T) \sqp^2
\Bigg\}. \label{free theory} \ena Here we have extracted from
$m^2$, $\kappa$, and $\nu$ the factors merely induced by the
rescaling. The influence of the environment on the spectrum of the
long wave modes will manifest through these terms
\cite{Starobinsky,Lombardo:Nacir, Matarrese, Liguori}.

The flow of the RG drives the initial interacting theory
(\ref{original theory}) towards the free theory (\ref{free
theory}), and allow us to find a relation between expectation
values associated with each theory. The relation is \bea
\label{relation} G\Big(k, t,\mu(\Lambda, T)\Big) =
\lt(\Lambda/k\rt)^{\alpha(k, T)} \; G\lt(\Lambda,\,
\lt(\Lambda/k\rt)^{\beta(k, T)} \, t, \mu(k, T)\rt). \ena On the
left hand side, $G$ is the two field expectation value computed
for a mode $k$ at time  $t$, and $\mu(\Lambda, T)$ stands for the
set of parameters which define the action at scale $\Lambda$. In
our case the only parameter is the coupling constant $\lambda$. On
the right hand side, $G$ is the expectation value of the theory
defined by the set of parameters $\mu(k, T)$, reached after modes
between $k$ and $\Lambda$ have been eliminated. The relevant
parameters in $\mu(k, T)$ are $m^2(k, T)$, $\kappa(k, T)$, and
$\nu(k, T)$. Finally, the exponents $\alpha$ and $\beta$ depend on
the trajectory followed by the action when it goes from scale
$\Lambda$ to $k$.

Now we connect to the original problem for the power spectrum of
an interacting inflaton field. We must feed the RG group equations
with an initial condition at scale $\Lambda$ and then use the
relation (\ref{relation}) to obtain the expectation value for the
mode $k$ as it exits the horizon. The initial condition, in terms
of the conformal field, is given by the CTP action at scale
$\Lambda$, Eq.(\ref{original theory}), where $t$ has to be
substituted by the conformal time $\eta$. According to
Eq.(\ref{eta exit}), the mode $k$ exits the horizon when \bea \eta
= -k^{-1}. \ena If inflation starts at $\eta^*$, the time that the
mode $k$ spends inside the horizon is given by \bea \tau_k =
-k^{-1} - \eta^*. \ena For the interacting theory
Eq.(\ref{espectro 0}), reads \bea \lt<\Phi(k,t) \Phi(k,t)\rt>_{\rm
HE} &= k^{-2} \; G\Big(k, \tau_k,\lambda\Big). \ena From
Eq.(\ref{relation}), identifying $t$ and $T$  with $\tau_k$, we
get \bea \fl \lt<\Phi(k,t) \Phi(k,t)\rt>_{\rm HE} &= k^{-2}
\lt(\Lambda/k\rt)^{\alpha(k, \tau_k)} \nn
\\ &\;\;\;\;\;\;\;\;\;\;\;\times G\lt(\Lambda, \,
\lt(\Lambda/k\rt)^{\beta(k, \tau_k)} \, \tau_k, \lt\{m^2(k,
\tau_k), \kappa(k, \tau_k), \nu(k, \tau_k) \rt\}\rt).
\label{relation 2} \ena Here, the relevant elements of $\mu(k,
\tau_k)$ have been shown explicitly. The right hand side of
Eq.(\ref{relation 2}) can be calculated using the $G$
corresponding to the action (\ref{free theory}) \bea \fl G\lt(k,
t, \lt\{m^2, \kappa, \nu\rt\}\rt) = \lt(\D\frac{2}{k} -
\D\frac{\nu}{\kappa \omega_0^2} \rt)
\lt[\D\frac{\omega_0^2}{\omega^2}  - \D\frac{\kappa^2}{\omega^2}
\cos(2\omega t) + \D\frac{\kappa}\omega \sin(2\omega t) \rt] e^{-2
\kappa t} + \D\frac{\nu}{\kappa \omega_0^2} \label{prop cG},\ena
where $\omega_0^2 = m^2+ k^2$ and $\omega^2 = \omega_0^2-
\kappa^2$  \cite{us}.

The expressions for $m^2$, $\kappa$, and $\nu$, and for the
exponents $\alpha$ and $\beta$ in Eq. (\ref{relation 2}), as
functions of $k$ and $\tau_k$ are given in \cite{us}. In Fig. 1 we
show $n(k)-1$,  defined in (\ref{spectral index}),  as function of
$k$ for a particular choice of $\lambda$ and $\eta^*$. (We have
chosen $-\eta^* = 200$, large enough so the ratio $k_{\rm max}/
k_{\rm min}$ can be about 100, the expected ratio between the
maximum and minimum length scales of the inhomogeneities.) The
main effects are introduced by the mass term.

\begin{figure}
\begin{center}
\includegraphics[width=11cm]{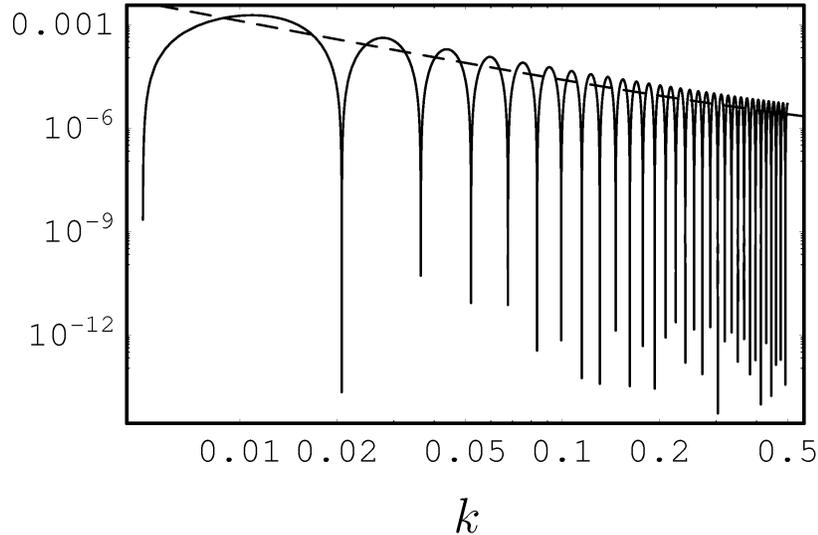}
\end{center}
\caption{$n(k)-1$ as function of $k$ (solid curve), with $\eta^* =
-200$, and $\lambda = 10^{-3}$. The dashed curve shows $n(k)-1$
once the rapidly oscillatory terms have been removed. The main
departures from zero come from the mass term induced by the coarse
graining.}
\end{figure}

We have presented a toy model for computing the spectrum of the
fluctuations of the inflaton field. At the present stage we do not
pretend to derive quantitative conclusions from this model. Our
main concern was to show that the same arguments usually given in
the context of the RG in equilibrium, can be extended to
nonequilibrium problems using the CTP formalism, where noise and
dissipation show up naturally.

\ack We acknowledge Joan Sola and  Enric Verdaguer for their kind
hospitality at Barcelona, and Enric Verdaguer for discussions.
This work was supported by Universidad de Buenos Aires, CONICET
and ANPCYT.

\end{document}